\documentclass[pre,twocolumn,showpacs,superscriptaddress,floatfix,reprint]{revtex4-1}

\usepackage[utf8]{inputenc}
\usepackage[english]{babel}
\usepackage[T1]{fontenc}
\usepackage{graphicx}
\usepackage{amsmath}
\usepackage{amsfonts}
\usepackage{amssymb}
\usepackage{siunitx}
\usepackage{textcomp}
\usepackage{standalone}
\usepackage{xspace}
\usepackage[caption=false]{subfig}
\usepackage[normalem]{ulem}
\usepackage{color}
\graphicspath{{figures/}} 
\begin{document}

\title{Study on the Kinetics of Rayleigh Particle Jets Converging by Laser Beams}
\author{Z. L. Wang}
\altaffiliation[Corresponding author: ]{wng\_zh@shu.edu.cn}
\affiliation{Shanghai Institute of Applied Mathematics and Mechanics, Shanghai Key Laboratory of Mechanics in Energy Engineering, Shanghai University, Yanchang Road 149, Shanghai, 200072, P.R. China}
\author{Nu Zhenyu}
\affiliation{Shanghai Institute of Applied Mathematics and Mechanics, Shanghai Key Laboratory of Mechanics in Energy Engineering, Shanghai
University, Yanchang Road 149, Shanghai, 200072, P.R. China}
\author{Huang Kai}
\affiliation{Shanghai Institute of Applied Mathematics and Mechanics, Shanghai Key Laboratory of Mechanics in Energy Engineering, Shanghai University, Yanchang Road 149, Shanghai, 200072, P.R. China}

\begin{abstract}
This paper discusses laser-induced flow stabilizing of Rayleigh particle jets. Laser technology, has important applications in micro/nano-scale static monomer particle operations, such as optical tweezers, or is used for the passive measurement of macroscopic physical features of particle groups. However, it is relatively rare for the laser beam to directly interfere with the behavior of particle populations dynamically, so as to achieve the purpose of instant group manipulations. Based on the theoretical analysis of particle dynamics and hydrodynamic stability theory, the effects of light induced convergence on rarified jets (as a point source emitting particle off a nozzle) and denser jets consists of Rayleigh sized particles have been considered. For rarified particle jet's analysis, compared with the classical vacuum evaporation deposition theory, we found that the laser positively guided the movement of particles, leading their pathes into more concentrated targets. Such convergence effect also happens in the case of denser Rayleigh particle jets. Particle dynamics simulations and hydrodynamic stability analysis mutually authenticated that the optical field forces suppress the instability both of long-wave and short-wave on particle jet interfaces, and have broad-spectrum stabilization characteristics. Therefor, diffusive particles in vacuum evaporation can also have very good targeted aggregations by laser.
\end{abstract}


\maketitle
\section{Introduction}
Jets composed of particulate matter are widely present in a variety of natural phenomena and engineering processes, such as inkjet printing, shot peening, pulverized coal gasification, and the like. In order to study particle jets, many methods for producing particle jets have been developed. Thoroddsen et al. used large-diameter ($1.34 cm$) high-velocity balls to impinge on a substrate packed with $80 \mu m$ diameter glass beads, demonstrating that the particulate matters can still generate high-density jets at the micro-scale through deformation-rebounding of the substrate~\cite{Thoroddsen2001}. Particle jets can also be formed by carrying particles through a high-speed air stream, and a liquid film-like flow structure can also be obtained after striking the wall surface. Shi et al. used such method to find that the thickness of the liquid film was independent of the flow velocity and particle size, and the evolution of the liquid film was related to the ratio of flow diameter and particle size~\cite{Shi2017}.

The demands for MEMS researches have been increasing in recent years. Researchers at Harvard University adopted $100 nm$ diameter powder particles in 2013 to successfully manufacture lithium-ion micro-cells with local accuracy of up to $1 \mu m$ by using 3D printing technology~\cite{Sun2013}. The MIT Microsystems Technology Laboratory developed a low-cost processing method based on electrospray printing technology in 2015, using a solution containing nanoparticles to produce a microsensor with only $0.03 mm^2$ large, whose metal wire of its internal circuit is only $10 \mu m$ wide with a minimum spacing of $50 \mu m$~\cite{Taylor2015}. Additive Manufacturing technology at a micro/nano scale has also been rapidly developed. Fuller et al. used piezoelectric ink-jet printing technology to transport nano-particles to successfully construct a variety of micro-electronic components, demonstrating the great potential of inkjet printing technology in the MEMS field~\cite{Fuller2002}. As an alternative to inkjet printing technology, Huang et al. used aerosol printing technology to successfully complete the low-cost and fast printing of conductive silver films on $A4$ paper substrates at room temperature~\cite{Huang2014}.

For the transportation of micro/nanoparticles in jet researches, the studies usually focus on particles of size no smaller than $1 \mu m$, and the influence of the fluid environment is often negligible. While for nanoparticles, there are many applications other than inkjet printing and aerosol printing. Xiao et al. used electrospinning technology to disperse nano-metal particles into fiber materials, which reduced the occurrence of particle agglomeration and improved the material properties~\cite{Xiao2009}. Hua et al. studied the effects of process parameters on the properties of aluminum films fabricated on single crystal silicon by DC magnetron sputtering, and found that the films prepared have preferred orientation for specific crystal structure particles~\cite{Hua2015}. Yu et al. successfully prepared high-purity, high-catalytic-efficiency SiC nanoparticles using heat plasma technology~\cite{Yu2017}.

Because particles transportation in jets often requires various liquids, gases, or other substances as carriers, the roles of carriers can not be ignored, which limits their practical applications. In this paper, we attempts to construct a theoritical basis for a direct writing method with little or non carrier effects by using the optical field confinement effect of the optical tweezer technique.

The study of optical tweezers originated in the $1970s$ when Ashkin in the laboratory accidentally observed the phenomenon of laser disturbance of colloidal particles in a solution~\cite{Ashkin1970a}. Later, Ashkin and his collaborators conducted theoretical experiments using lasers for atom/molecule and particle capturing~\cite{Ashkin1970b, Ashkin1978}. Until 1986, Ashkin successfully used a single-beam gradient force trap for the first time in the laboratory to achieve stable capture of colloidal particles ranging from $25 nm$ to $10 \mu m$, which marked the official birth of optical tweezers~\cite{Ashkin1986}. Optical tweezers technology has been widely used in microscale researches such as chromosome separation and measurement of particle-particle interactions~\cite{Ren2008, Wang2004}.

The existing optical technology literatures focus on the researches of manipulations of single particles. The research on the influence of a large number of particles on population behavior is rare in literatures and patents. While studying the vacuum evaporation coating technology of organic light-emitting diodes (OLEDs), inspired by the phenomenon of light trapping particles near the optical axis, this article attempts to explore the feasibility of using optical tweezers to control the particulate flow converging or focusing~\cite{HuangKai2013}. The force calculation method for spherical particles in an electromagnetic field had been given in the work of Lorenz, Mie and Debye, Lorenz-Mie theory, but their method are cumbersome to calculate. After the birth of optical tweezers technology, in order to meet the needs of optical tweezer designs, people had developed a variety of methods to simplify the calculation processes. The methods applied to the light force of the particles computing is generally differentiated according to $\lambda/a$, the ratio of the laser wavelength to the radius of the particles. When $\lambda/a>20$, the particles are called Rayleigh particles, and very accurate results can be obtained using the electric dipole model~\cite{Harada1996}. And when $\lambda/a<1/5$, particles, called Mie particles, can be approximated by geometrical optics~\cite{Ashkin1970a}. The more generalized GLMT (Generalized Lorenz-Mie theory) method can be applied to the calculation of spherical particle forces at various scales~\cite{Gouesbet1988}.

In our previous work, the calculations of Mie particles were testified by means of theoretical analysis and numerical simulation~\cite{HuangKai2013m, WangHan2016m}. The discussions in this article will focus on the behavior of nano-sized Rayleigh particles under laser light beams.

\begin{figure}\centering
    \subfloat[(a)]{\includegraphics[height=4cm]{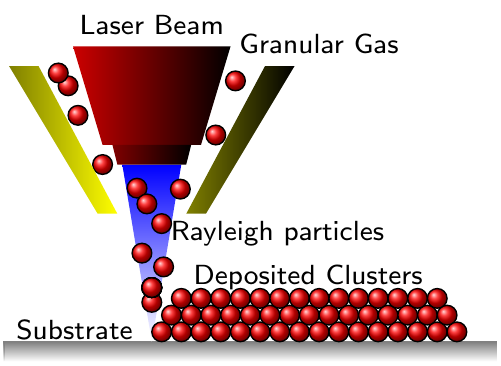}\label{fig1a}}
    \\
    \subfloat[(b)]{\includegraphics[height=6cm]{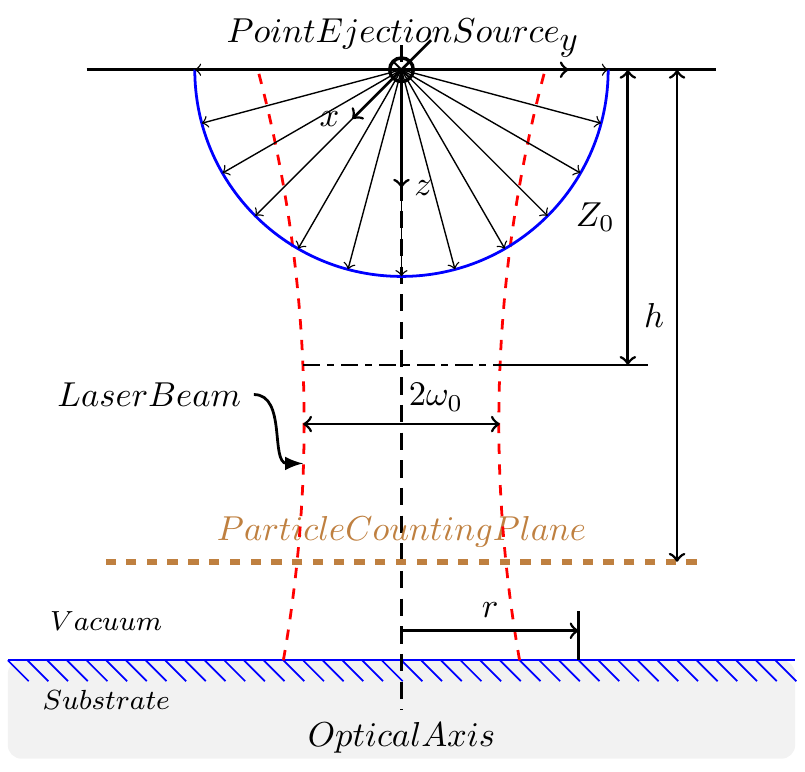}\label{fig1b}}
    \caption{ \protect\subref{fig1a} Conceptual prototype of laser induced deposition.  \protect\subref{fig1b}. Simplified computational model as particles emitting from a point source into a Gauss laser beam.}
\end{figure}

\section{Physical model and simulation methods}
\subsection{Physical model}

In this paper, the software LAMMPS(Large-scale Atomic/Molecular Massively Parallel Simulator) was used for numerical simulation. Lammps is a molecular dynamics software developed by the Sandia National Laboratories in the United States. Modeling simulations of motions in different force fields and boundary conditions such as atoms, biomolecules, particle streams, or other coarse-grained systems can be performed. At the same time, the MPI (Message Passing Interface) is integrated into its software architecture, simplifying the implementation of parallel computing ~\cite{Plimpton1995Fast}. And the software can be sourced from its official website (http://lammps.sandia.gov) and authorized to be modified when necessary.

The initial physical model can be seen in Figure 1(a) as a combination of Knudson box (particle source) and deposition liner used in vacuum evaporation. The laser is added to emit from the particle outflow hole, pointing to the liner or collection device, and inducing the movement of molecular clusters or particles in a vacuum environment.

After initially attempting to use LAMMPS to simulate the position of the deposition liner from the inside of the evaporation chamber to the outside, it was found that the flow simulation inside of Knudson box requires a large amount of computational resources. And although the Knudsen number (average molecular free path to orifice diameter ratio) is at a minimum of $0.003$, which is close to a viscous flow state. However, the particle flow outflow and velocity distribution are still affected by parameters such as pore length, mean free path and pore size, and they behave more like molecular flows. And there are too many variables that affect the simulation results~\cite{Masayoshi2009}. As the ultimate concern is the convergence effect of the laser on the particles, taking considerations comprehensively, the model can be simplified as combinations of a particle source and the external laser field. The simplified model is shown in Figure 1(b). Where $r$ is the distance of a point on the counting plane to the optical axis; $\omega_0$ is the waist radius of the laser; $h$ is the coordinate of the counting plane in the $z$ direction; $z_0$ is the $z$ coordinate of the waist of the laser beam.

This article will select the point source and the surface source, respectively, to discuss the effects of laser beams on the convergence of the Rayleigh particle jet flow.

\subsection{Rayleigh particles affected by laser force}
According to Harada's work, the scattering force $F_{scat}$ and the gradient force $F_{grad}$ produced by the laser beam on Rayleigh particles can be calculated by Equation (1) ~\cite{Harada1996}.

\begin{equation}
\left\{
\begin{aligned}
{{\bf{F}}_{{\rm{scat}}}} &= \hat z\frac{8}{3}\pi {k^4}{a^6}\frac{{{m^2} - 1}}{{{m^2} + 2}}\frac{n}{c}I\left( {x,y,z} \right)\\
 {{\bf{F}}_{{\rm{grad}}}} &= \frac{{2\pi n{a^3}}}{c}\left( {\frac{{{m^2} - 1}}{{{m^2} + 2}}} \right)\nabla I(x,y,z)\end{aligned}
\right.
\end{equation}

Where $\hat z$ is a unit vector in the $z$ direction, which is also the laser propagation direction; the wave number $k = 2\pi/\lambda$, where $\lambda$ is the laser wavelength; $a$ is the radius of the particle; $m$ is the relative refractive index of the particle to the environment; $n$ is the refractive index of the medium in which the particles are located; $c$ is the speed of light; and $I(x,y,z)$ is the average Poynting vector.

In this paper, the common linearly polarized Gaussian beam is selected as the laser source. The average Poynting vector is written as Equation (2) in the coordinate system in Fig. 1(b).

\begin{equation}
I\left( {x,y,z} \right) = \frac{{\left( {\frac{{2P}}{{\pi \omega _0^2}}} \right)}}{{1 + {{\left( {2\frac{{z - {z_0}}}{{k\omega _0^2}}} \right)}^2}}}Exp[\frac{{ - 2\frac{{{x^2} + {y^2}}}{{\omega _0^2}}}}{{1 + {{\left( {2\frac{{z - {z_0}}}{{k\omega _0^2}}} \right)}^2}}}]
\end{equation}

\subsection{Numerical aspects}
The size of the simulation box set here is $100 \mu m\times 100 \mu m\times 81 \mu m$. Different particle sources are placed at the origin of the coordinates of Figure 1(b), emitting particles in the positive $z$-axis direction.

Rayleigh particles here were considered equivalent to ideal spheres in our simulations, without introducing extra complexity of their specific morphology. Particles move in an almost vacuum environment. Therefore, the main considerations are the interaction of particles with each other, and also the interactions between particle and laser beam.

According to the setting scale of the model, this paper deals with a mesoscale problem. The commonly used particle methods on such scales include multi-particle collision dynamics (MPC), dissipative particle dynamics (DPD), and direct simulation Monte Carlo (DSMC), etc.~\cite{Noguchi2007}. The DPD method is often used to study the movement of macromolecules in fluids. By applying coarse-grained modeling of fluids and molecules to the applied potential obtained by statistical methods, the amount of calculations can be reduced to speed up the simulation process. The coarse-grained process makes the potential of the DPD method naturally include a part that reflects the viscous and thermal motion of the fluid environment~\cite{Groot1997}. MPC and DSMC methods are similar to DPD, have force terms counting the influence of the fluid environment. However, the objects that these methods target are different from the goals that this article is trying to simulate.

Referring to the work of Israelachvili, ~\cite{Israelachvili1985}, the attractive force between nano-sized homogeneous spherical particles decays rapidly within a surface spacing much smaller than the diameter, and it attracts each other during extremely close distances and collisions. It is difficult to calculate due to the deformation of the particles, and the overall repulsive force manifests itself in the elasto-plastic deformation of the particles when they collide. At present, no description has been found for the particle attractiveness that fits this problem. According to Hamaker's work, the mutual attraction between the particles in the simulation is simulated, and the attractive force can approach the laser force when the particle surface spacing is smaller than the diameter ~\cite{Hamaker1937}. For the repulsion between nanoparticles, considering the particle flow study, due to the observation of particles larger than $1 \mu m$, the mutual attraction between particles due to Van der Waals forces can be ignored. The particle collision process is calculated according to the extrusion deformation, in which the plastic deformation occurs. The impact of mass changes, etc., is reflected in collisional dissipation ~\cite{Goldhirsch2003}. Therefore, in this paper, the impact of particle flow collision developed by Silbert and Brilliantov was used in the simulation to simulate the repulsion between nanoparticles ~\cite{Silbert2002Geometry, Brilliantov1996}, but the weak mutual attraction between nanoparticles was not simulated. In this paper, the simulation of the interaction between nanoparticles is not perfect, and further work is needed for discussion.

At this time, the interaction between the spherical particles i,j is calculated according to the collisional effect of the particle stream, ie, equation (3).

\begin{equation}
{{\bf{F}}_{{\rm{ij}}}} = \left( {{k_n}\delta {n_{ij}} - {m_{eff}}{\gamma _n}{v_n}} \right) - ({k_t}{\rm{\Delta }}{s_t} + {m_{eff}}{\gamma _s}{v_s})
\end{equation}

Where $k_n$ is the normal elastic coefficient between the particles (ball center line direction); $\delta n_{ij}$ is the variation of the center distance of the particle; $m_{eff}=m_i m_j/(m_i+m_j)$ is the effective mass at the time of particle collision; $\gamma_n$ is the normal damping Coefficient; $v_n$ is the component of the relative speed of two particles in the normal direction; $k_t$ is the tangential elasticity coefficient; $\Delta s_t$ is the tangential displacement vector of two balls; $\gamma_s$ tangential damping coefficient; $v_s$ is the tangential component of relative velocity. Among them, the elastic coefficient and the damping coefficient need to be manually set. Since the effects of collisional dissipation and particle rotation are not considered in this paper, $k_t=0, k_n=0.005 nN/\mu m$ are set in the simulation, and the damping coefficient calculation is turned off.

In the simulation, the particle motion trajectory is obtained by integrating time (4) and equation (5), that is, the Velocity-Verlet algorithm ~\cite{Frenkel1997}.

\begin{equation}
{\bf{r}}\left( {t + {\rm{\Delta }}t} \right) = {\bf{r}}\left( t \right) + {\bf{v}}(t){\rm{\Delta }}t + \frac{{{\bf{F}}\left( t \right)}}{{2m}}{\rm{\Delta }}{t^2}
\end{equation}

\begin{equation}
{\bf{v}}\left( {t + {\rm{\Delta }}t} \right) = {\bf{v}}\left( t \right) + \frac{{{\bf{F}}\left( {t + {\rm{\Delta }}t} \right) + {\bf{F}}\left( t \right)}}{{2m}}{\rm{\Delta }}t
\end{equation}

Where $r$ is the particle position vector, $v$ is the particle velocity vector, $F$ is the particle force, $t$ is the current time, and $\Delta t$ is the time step.

In this paper, glycerol, a common test substance, was selected as the source of particle parameters for numerical simulation, and the environmental assumption of vacuum evaporation was used. The particle density $\rho=1.261×10^3 kg/m^3$, the relative refractive index $m=1.4746$, and the particle radius a is $0.03 \mu m$. Ambient refractive index $n=1$(vacuum).

\section{Different particle source simulations and results}
\subsection{Point source simulation}
This article first uses the point source model to begin research, preliminary verification procedures and the feasibility of laser beam collection particles. The initial velocity of the particles needs to satisfy a certain velocity distribution. For this reason, reference is made herein to the description of various evaporation sources in the vacuum evaporation technique. In many vacuum evaporation processes, particles do not almost collide during the flight except at the outlet of the evaporation source ~\cite{Ohring2001}. This article chooses a point source that is easy to get the velocity distribution as a starting point. The particle velocity probability density distribution from the point particle source is isotropic. Based on this property and the basic principles of statistical mechanics, it is possible to give a system of equations that satisfies the probability density function $f(v)$ of the velocity vector v in the polar coordinate space when the particle mean initial kinetic energy $\overline{\epsilon}$ is constant. 6) Further, the probability density distribution of the initial velocity of the particle originated from the point source is derived as formula (7), and the probability density distribution satisfying each velocity component is given by (8) ~\cite{ZhaoKaihua2005}.
\begin{equation}
\left\{
\begin{aligned}
\mathop \smallint \limits_0^{2\pi } {\rm{d}}\phi \mathop \smallint \limits_0^\pi  \sin \theta {\rm{d}}\theta \mathop \smallint \limits_0^\infty  {v^2}{\rm{d}}vf(v,\theta ,\phi ) &= 1\\
 \frac{1}{2}m\mathop \smallint \limits_0^{2\pi } {\rm{d}}\phi \mathop \smallint \limits_0^\pi  \sin \theta {\rm{d}}\theta \mathop \smallint \limits_0^\infty  {v^4}{\rm{d}}vf(v,\theta ,\phi ) &=\overline{\epsilon}
 \end{aligned}
\right.
\end{equation}
\begin{equation}
{\rm{f}}\left( v \right) = {\left( {\frac{{3m}}{{4\pi \overline{\epsilon} }}} \right)^{\frac{3}{2}}}Exp\left[ { - \frac{{3m{v^2}}}{{4\overline{\epsilon} }}} \right]
\end{equation}
\begin{equation}
{{\rm{f}}_i}\left( {{v_i}} \right) = {\left( {\frac{{3m}}{{4\pi \overline{\epsilon} }}} \right)^{\frac{1}{2}}}Exp\left[ { - \frac{{3mv_i^2}}{{4\overline{\epsilon} }}} \right],(i = x,y,z).
\end{equation}

With reference to Oring's work on vacuum deposition deposition thickness, it can be obtained on the counting plane shown in Fig. 1(b). When no laser beam is collected, the density of the number of particles passing through the plane at the distance $r$ from the optical axis is satisfied. Dimensionless Distributed (9) ~\cite{Ohring2001}.

\begin{equation}
\frac{{\rm{N}}}{{{{\rm{N}}_0}}} = \frac{1}{{{{\left( {1 + {{\left( {\frac{r}{h}} \right)}^2}} \right)}^{\frac{3}{2}}}}}.
\end{equation}
Where $N$ is the number of crossings of the particle from the optical axis $r$; the number of crossings of the particle at the $N_0$ optical axis.

$300,000$ particles were randomly injected into the calculation area from the origin of the coordinates in a time range of $0.42 s$. The particle velocity is assigned according to the velocity distribution (8) when entering the simulation zone. Although the point source emission particles are omnidirectional, only the particles moving upward hemisphere as shown in Fig. 1(b) are simulated here. In addition, the lower limit of the emission distance should be set during the simulation to ensure that the newly injected particles will not collide with the already-launched particles at the time of injection, so that the energy cannot be conserved during the collision calculation due to the excessive distance. In order to reduce the number of simulation steps, the allowable time step $\Delta t$ is automatically adjusted in the course of the simulation according to the maximum velocity of the particles and the defined single-step maximum movement distance in the interval of $0.0005 \mu s$ to $0.05 \mu s$.

Point particle source simulation laser wavelength $\lambda = 632.8 nm$, beam waist position $z_0 = 5 \mu m$, waist waist radius $\omega_0 = 5 \mu m$. The average initial kinetic energy of point particle source simulation is $\overline{\epsilon}= 0.0388 eV$.

The cross-over density distribution at $h=10 \mu m$ under the action of different power lasers is shown in FIG. 2 . Fig. 2(a) is the original calculation result of crossing density, and Fig. 2(b) is the result of dimensionless reference formula (9).

\begin{figure*}\centering
    \subfloat[(a)]{\includegraphics[height=6cm]{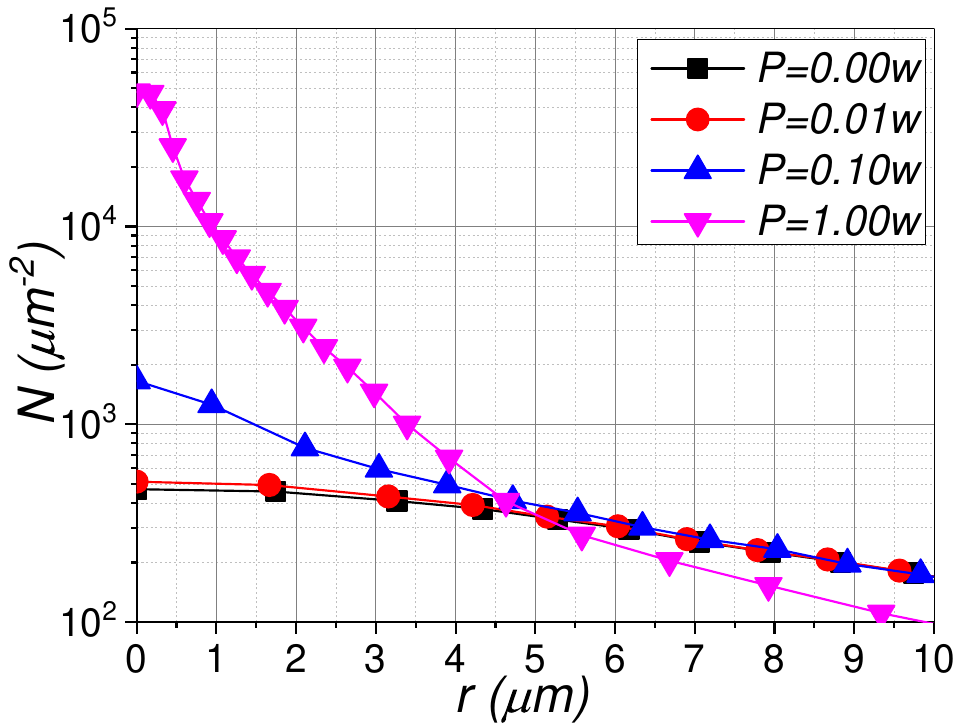}\label{fig2a}}
    \subfloat[(b)]{\includegraphics[height=6cm]{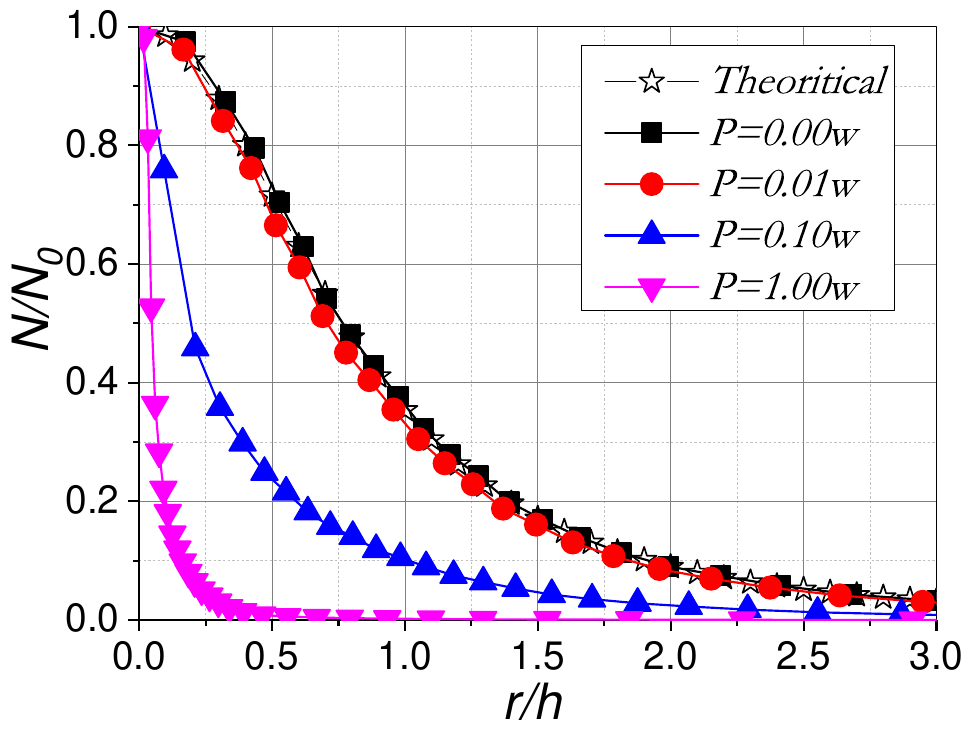}\label{fig2b}}
    \caption{ Comparison of the distribution and theoretical distribution of the crossing density at the counting plane $h=10$ under different power influences. \protect\subref{fig2a}. Density distribution at different powers.  \protect\subref{fig2b}. Non-dimensionalized distribution.}
\end{figure*}

From FIG. 2 , it can be seen that as the laser power increases, the distribution of the particles is concentrated near the optical axis of the laser ($r=0$). In the absence of a laser effect ($P=0.00W$), Fig. 2(b) shows that the simulation results are returned to the dimensionless distribution form that should be followed in the case of no bundles calculated by equation (9). Therefore, for the point source, the laser can effectively converge the trajectory of the particles in the space under certain parameters, improve the distribution in the specified plane, and at the same time, there is no abnormality in the movement of the particles after a few collisions in the simulation process. This program can simulate the collision and movement of particles normally.

\subsection{Theoretical analysis and simulation of light converging nanoparticle jets}
\subsubsection{Light Convergence Particle Jet Stability}

Lun gave the control equation (10) of the hydrodynamic form of the particle flow in 2006 ~\cite{Lun2006}.
\begin{equation}
\left\{
\begin{aligned}
\frac{{\rm{D}}}{{{\rm{D}}t}}\rho  &=  - \rho \nabla  \bullet {\bf{V}}\\
 \rho \frac{{{\rm{D}}{\bf{V}}}}{{{\rm{D}}t}} &= \rho {\bf{F}} - \nabla  \bullet {\bf{P}}\\
 \frac{3}{2}\rho \frac{{{\rm{DT}}}}{{{\rm{D}}t}} &=  - \nabla  \bullet Q - P:\nabla V - \Gamma
 \end{aligned}
\right.
\end{equation}

Where $\rho$ is the local average mass density, $T$ is the particle temperature, $Q$ is the heat flow vector of the particle stream, $P$ is the pressure tensor of the particle stream, $\Gamma$ is the energy dissipation rate per unit volume, and $F$ is the unit mass of the acceptor force.
The particle temperature $T$ in the particle stream does not refer to the thermodynamic temperature of the material and is the average measure of the pulsation velocity. For specific calculations, see equation (11).
\begin{equation}
\left\{
\begin{aligned}
n\left( {r,t} \right) &= \mathop \smallint \nolimits f\left( {v,r,t} \right)dv\\
V\left( {r,t} \right) &= \frac{1}{{n\left( {r,t} \right)}}\mathop \smallint \nolimits vf\left( {v,r,t} \right)dv\\
{\rm{T}}\left( {r,t} \right) &= \frac{1}{{n\left( {r,t} \right)}}\mathop \smallint \nolimits {\left( {v - V} \right)^2}f\left( {v,r,t} \right)dv
 \end{aligned}
\right.
\end{equation}

Where $f(v,r,t)$ is the particle velocity distribution function, $n(r,t)$ is the number density of particles, and $V(r,t)$ is the velocity field distribution.

The $Q$, $P$, and $\Gamma$ of the Lun structure are all functions of the volume fraction $\nu=\rho/\rho_p$ (ratio of the mass density of the jet to the particle density) and the particle temperature. The specific form is relatively complex and can be found in the corresponding literature~\cite{Lun2006}.

In previous work, the action of the laser was introduced by the volume force term $F$, ie equation (12).

\begin{equation}
\left\{
\begin{aligned}
{\bf{F}} &=  - \nabla  \bullet \mathord{\buildrel{\lower3pt\hbox{$\scriptscriptstyle\leftrightarrow$}}
\over {\cal T}}
- \frac{{\partial \mathord{\buildrel{\lower3pt\hbox{$\scriptscriptstyle\rightharpoonup$}}
\over g} }}{{\partial t}}\\
\mathord{\buildrel{\lower3pt\hbox{$\scriptscriptstyle\leftrightarrow$}}
\over {\cal T}}  &=  - \left( {{\overline{\epsilon} _0}EE + \frac{1}{{{{\rm{\mu }}_0}}}BB} \right) + \frac{1}{2}({\overline{\epsilon} _0}{{\rm{E}}^2} + \frac{1}{{{{\rm{\mu }}_0}}}{{\rm{B}}^2})\mathord{\buildrel{\lower3pt\hbox{$\scriptscriptstyle\leftrightarrow$}}
\over j } \\
\mathord{\buildrel{\lower3pt\hbox{$\scriptscriptstyle\rightharpoonup$}}
\over g}  &= {\overline{\epsilon} _0}E \times B
 \end{aligned}
\right.
\end{equation}

Where $\mathord{\buildrel{\lower3pt\hbox{$\scriptscriptstyle\leftrightarrow$}}
\over {\cal T}} $ is the momentum flow density of the electromagnetic field and $\mathord{\buildrel{\lower3pt\hbox{$\scriptscriptstyle\rightharpoonup$}}
\over g}$ is the momentum density of the electromagnetic field ~\cite{HuangKai2013m}. $E$ and $B$ are the electric field and magnetic field vectors, respectively. $\mathord{\buildrel{\lower3pt\hbox{$\scriptscriptstyle\leftrightarrow$}}
\over j }$ is a unit tensor.

The following basic flow was used for stability analysis, where density $\bar \rho  = {\rho _0} = const$, pressure distribution $\bar p = \bar p(r)$, temperature distribution $\bar T = \bar T(r)$, outlet jet radius $\bar h = {R_0}$, outlet flow rate ${\bf{\bar V}} = (0,0,{\bar V_{\rm{z}}} = U = const)$, the free surface boundary radius ${R_0}$ at the outlet, the outlet volume fraction $\bar \nu  = {\nu _0} = {\rho _0}/{\rho _p} = const$, and ${\rho _p}$ is the particle density. Then superimpose the small perturbation on the elementary flow and substitute the non-dimensionalized particle flow control equation to obtain the perturbation control equation (13).

\begin{widetext}
\begin{eqnarray}
\left\{
\begin{aligned}
\frac{{\partial \hat \rho }}{{\partial \hat t}} + \frac{{\partial \hat \rho }}{{\partial \hat z}} &=  - \hat \nabla  \bullet {\bf{\hat V}}\\
\frac{{\partial {{\hat V}_z}}}{{\partial \hat t}} + \frac{{\partial {{\hat V}_z}}}{{\partial \hat z}} &= {\hat a_F}{\hat {\bar F}_z}\hat \rho  - \frac{{\partial \left( {{{\hat a}_p}\hat \rho  + {{\hat b}_p}\hat T} \right)}}{{\partial \hat z}} + \frac{2}{{{\rm{Re}}}}\frac{\partial }{{\partial \hat z}}\frac{{\partial {{\hat V}_z}}}{{\partial \hat z}} + \frac{1}{{{\rm{Re}}}}\frac{1}{{\hat r}}\frac{\partial }{{\partial \hat r}}\hat r(\frac{{\partial {{\hat V}_z}}}{{\partial \hat t}} + \frac{{\partial {{\hat V}_r}}}{{\partial \hat z}}) + \frac{1}{{{\rm{R}}{{\rm{e}}_{\rm{\lambda }}}}}\frac{\partial }{{\partial \hat z}}\left( {\hat \nabla  \bullet {\bf{\hat V}}} \right)\\
\frac{{\partial {{\hat V}_r}}}{{\partial \hat t}} + \frac{{\partial {{\hat V}_r}}}{{\partial \hat z}} &= {\hat a_F}{\hat {\bar F}_r}\hat \rho  - \frac{{\partial \left( {{{\hat a}_p}\hat \rho  + {{\hat b}_p}\hat T} \right)}}{{\partial \hat r}} + \frac{1}{{{\rm{Re}}}}\frac{\partial }{{\partial \hat z}}\left( {\frac{{\partial {{\hat V}_z}}}{{\partial \hat t}} + \frac{{\partial {{\hat V}_r}}}{{\partial \hat z}}} \right) + \frac{2}{{{\rm{Re}}}}\frac{1}{{\hat r}}\frac{\partial }{{\partial \hat r}}(\hat r\frac{{\partial {{\hat V}_r}}}{{\partial \hat r}}) - \frac{1}{{{\rm{Re}}}}\frac{{{{\hat V}_r}}}{{{{\hat r}^2}}} + \frac{1}{{{\rm{R}}{{\rm{e}}_{\rm{\lambda }}}}}\frac{\partial }{{\partial \hat r}}\left( {\hat \nabla  \bullet {\bf{\hat V}}} \right)\\
\frac{{\partial \hat T}}{{\partial \hat t}} + \frac{{\partial \hat T}}{{\partial \hat z}} &= \frac{1}{{{\rm{RePr}}}}\left( {\frac{\partial }{{\partial \hat z}}\left( {\frac{{\partial \hat T}}{{\partial \hat z}}} \right) + \frac{1}{{\hat r}}\frac{\partial }{{\partial \hat r}}(\hat r\frac{{\partial \hat T}}{{\partial \hat r}})} \right) - {\sigma _T}\hat {\bar T}\left( {\hat \nabla  \bullet {\bf{\hat V}}} \right) + \left( {{{\hat a}_{\rm{\Gamma }}}\hat \rho  + {{\hat b}_{\rm{\Gamma }}}\hat T} \right)
\end{aligned}
\right.
\end{eqnarray}
\end{widetext}

The Reynolds number in the perturbation control equation ${\rm{Re}} = (U{R_0}{\rho _0})/{\rm{\bar \varsigma }}$, the additional Reynolds number ${\rm{R}}{{\rm{e}}_{\rm{\lambda }}} = (U{R_0}{\rho _0})/\bar \lambda $, the electro-Eula number ${{\rm{E}}_{\rm{u}}} = (0.25{\epsilon _0}E_0^2)/({\rho _0}{U^2})$, The Prandtl number ${\rm{Pr}} = 1.5{\rm{\bar \varsigma }}/\bar K$, where $\epsilon _0$ is the dielectric constant of the environment and $E_0$ is the maximum electric field field strength in the laser light field. Average particle viscosities ${\rm{\bar \varsigma }}$ and $\bar \lambda $, average thermal conductivity $\bar K$, and other coefficients ${\hat a_p}$, ${\hat b_p}$, ${\sigma _T}$, ${\hat a_{\rm{\Gamma }}}$, ${\hat a_F}$, are more complex and can be found in the literature ~\cite{HuangKai2013m}.

The solution (14) of the elementary stream temperature field ${\rm{\hat {\bar T}}}\left( {\hat r} \right)$ and the radial force ${\hat {\bar F}_r}\left( {\hat r} \right)$ can be obtained from equation (13), where the form of ${\hat a_t}$, ${\hat b_t}$, $\bar \Theta \left( r \right)$) and ${\Re ^{\left( 1 \right)}}(r)$ can be found in the corresponding literature ~\cite{HuangKai2013m}.

\begin{equation}
\left.
\begin{aligned}
{\rm\hat {\bar T}}\left( {\hat r} \right) &= \frac{{{{\hat b}_t}\hat {\bar F}{{\left( 1 \right)}^2}}}{{\hat r\hat {\bar F}{{\left( {\hat r} \right)}^2}\left( { - 1 + {{\hat b}_t}\hat {\bar F}{{\left( 1 \right)}^2}\mathop \smallint \nolimits_1^{\hat r}  - \frac{{2{{\hat a}_t}}}{{\hat {\bar F}{{\left( t \right)}^3}{t^2}}}{\rm{d}}t} \right)}}\\
{\hat {\bar F}_r}\left( {\hat r} \right) &= {E_u}\bar \epsilon {\hat a_0}\bar \Theta \left( r \right)[\frac{\partial }{{\partial \hat r}}[{\Re ^{\left( 1 \right)}}{\left( {\rm{r}} \right)^2}]]
 \end{aligned}
\right.
\end{equation}

The following stream of particles was then constructed for numerical analysis. Under a vacuum environment, the jet of particles exits from a nozzle with a radius ${R_0} = 3mm$ at a rate of $U = 100m/s$. The outlet volume fraction of the jet is $\nu  = 0.1$ and the maximum volume fraction $\nu _m = 0.55$. The diameter of the particles was $100nm$, the density was $1.5g/c{m^3}$, and the refractive index was $1.3$. Near-elastic collisions occur between the particles, and the elastic recovery coefficient $\eta  = 0.95$. The particle temperature of the jet is $10K$. Since the focus is on the influence of the laser on the particle jet, the inviscidity assumption (${\rm{Re}} , {\rm{R}}{{\rm{e}}_{\rm{\lambda }}} \to \infty $) is introduced to simplify the subsequent analysis.

Referring to Drazin's work ~\cite{Drazin1984}, use the regular modal method to write the disturbance variables ${\rm{\hat \rho }},{\hat V_z},{\hat V_r},{\rm{\hat T}}$ in the form of equation (15). Where ${\rm{\Omega }}$ is the time growth factor and $k$ is the axial wave number. The flow is stable when ${\rm{\Omega }<0}$, and the flow is neutrally stable when ${\rm{\Omega }=0}$, and the flow will destabilize with time when ${\rm{\Omega }>0}$.

\begin{equation}
\left\{ {\hat \rho ,{{\hat V}_z},{{\hat V}_r}, \hat T} \right\} = \{ \hat \rho (r),{\hat V_z}(r),{\hat V_r}(r),\hat T(r)\} exp({\rm{\Omega }}\hat t)exp(ik\hat z)
\end{equation}

It is substituted into the equation set (13) and simplified using the inviscid assumption to obtain a second-order ordinary differential equation.

\begin{equation}
\frac{{{\partial ^2}}}{{\partial {{\hat r}^2}}}\left( {\hat r{{\hat V}_r}} \right) + {A_1}\left( r \right)\frac{\partial }{{\partial \hat r}}\left( {\hat r{{\hat V}_r}} \right) + {A_0}\left( r \right)\left( {\hat r{{\hat V}_r}} \right) = 0
\end{equation}

where, ${A_0}\left( r \right)=({\rm{\Omega }} + ik)/{\rm{\Psi }}(r)$, ${A_1}\left( r \right)=-[{{\hat a}_F}{{\hat {\bar F}}_r}{B_\rho }\left( {\rm{r}} \right) - \frac{{\partial {\rm{\Psi }}\left( r \right)}}{{\partial \hat r}} + \frac{1}{{\hat r}}{\rm{\Psi }}\left( r \right)]/{\rm{\Psi }}\left( r \right)$, ${\rm{\Psi }}\left( r \right)=\left( {{{\hat a}_p} + {{\hat b}_p}{B_T}\left( r \right)} \right){B_\rho }(r)$, ${B_\rho }\left( r \right)=- {\rm{\Omega }} + ik + ik{B_{{V_z}}}{\left( r \right)^{ - 1}}$, ${B_{{V_z}}}\left( r \right) = \left( {{{\hat a}_F}{{\hat {\bar F}}_r} - ik{{\hat a}_p} - ik{{\hat b}_p}{B_T}\left( r \right)} \right)/({\rm{\Omega }} + ik)$, and ${B_T}\left( r \right)=\left( {{\sigma _T}\hat {\bar T}\left( {{\rm{\Omega }} + ik} \right) - {{\hat a}_{\rm{\Gamma }}}} \right)/({\rm{\Omega }} + ik + {{\hat b}_{\rm{\Gamma }}})$.

\begin{figure*}\centering
    \subfloat[(a)]{\includegraphics[height=6cm]{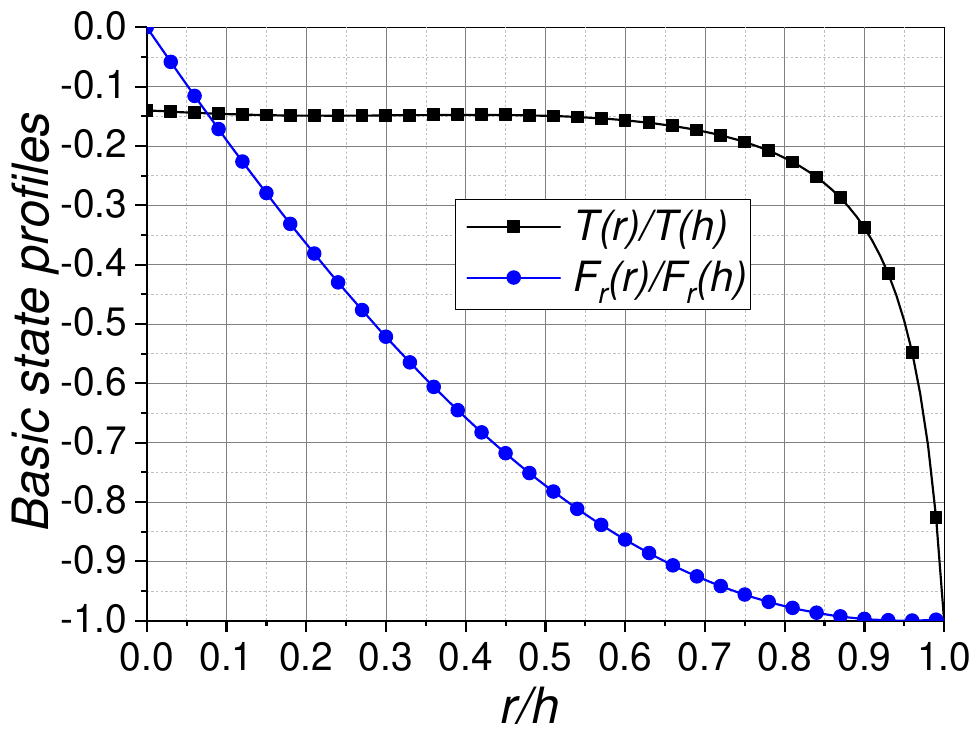}\label{fig3}}
    \subfloat[(b)]{\includegraphics[height=6cm]{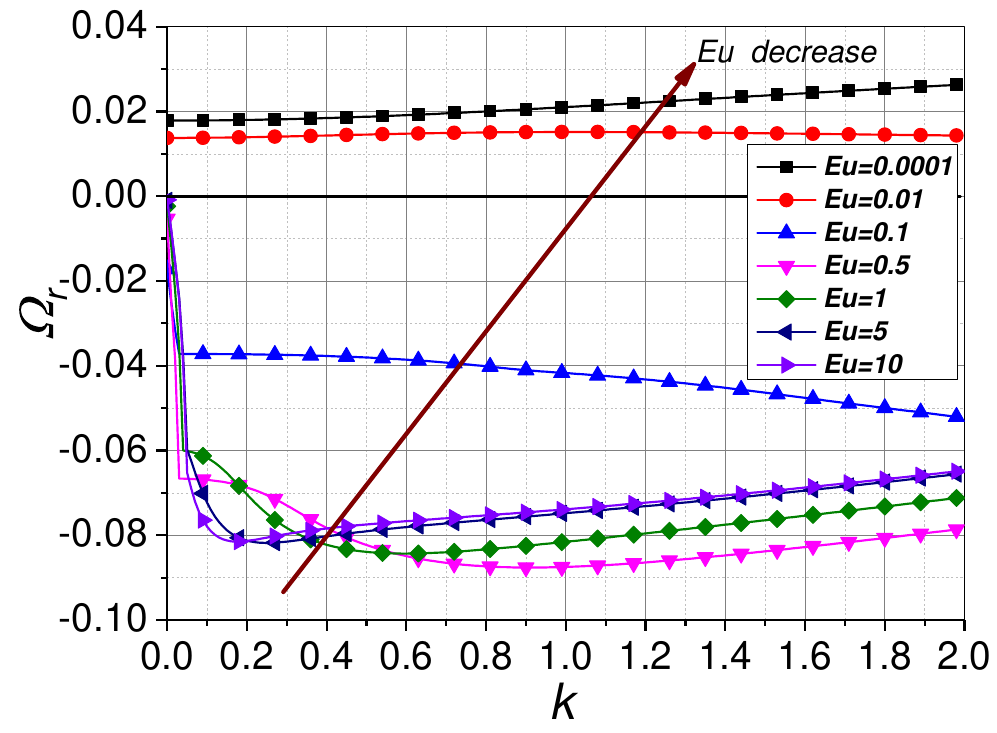}\label{fig4}}
    \caption{Hydrodynamic stability analysis of light forces on particle jets. \protect\subref{fig3} Normalized basic state radial force $\bar F_r$ and particle temperature $\bar T$.  \protect\subref{fig4} Effect of electro-Euler number $Eu$ on flow stability.}
\end{figure*}

Then, the new equation (16) is again transformed into a linear system of equations (17) using a pseudo-spectral configuration method for numerical solution. The system of equations is a homogeneous determinant, and the existence of a nontrivial solution is that the determinant of the matrix is zero.
\begin{equation}
\Xi \left( {{\rm{\Omega ,}}k;OtherPars} \right) = 0
\end{equation}

The equation set (17) was solved using the Müller method ~\cite{Moler1973}. Figure 3 reflects the calculated trends of radial force and particle temperature. In the figure, the radial force always points to the optical axis and increases with the increase of the radius, so that the particles converge toward the optical axis and the jets tend to converge. The analytical expression of $\bar F_r$ also shows that the laser power is increased and the binding force is increased. The distribution of particle temperature shows that the particle's activity in the vicinity of the boundary decreases rapidly while the internal change is relatively small. Fig. 4 is the relationship between the axial wave number $k$ and the growth factor ${{\rm{\Omega }}_{\rm{r}}}$ under different electro Euler numbers $E_u$. The electron Euler number $E_u$ decreases in the direction of the arrow in the figure, and the growth factor increases. From the situation change in the figure, it can be seen that, as the Euclidean number $E_u$ decreases as ${E_u} \le 0.01$, the growth factor ${{\rm{\Omega }}_{\rm{r}}}>0$, the analysis system according to the regular modal method will be unstable, and the stable jet cannot be maintained; and when the Eulerian When the number ${E_u} \ge 0.5$, the system remains stable. The most direct way to increase the Eulerian number $E_u$ is to increase the laser power, which is consistent with the concept of using the laser to stabilize the jet. Figure 4 also shows that the laser suppresses both the long-wave and short-wave instability of the jet. This is different from the typical surface force in the common stability theory—the selectivity of the surface tension on the disturbance wave. Usually the surface tension has a destabilizing effect on the long-wave surface of the jet, and it has an inhibitory effect on the short-wave.

\subsubsection{Particle Dynamics Simulation of Nanoparticle Jet}
In the real scientific research or production process, the particle source may be nozzles, helium and other components, and have a certain export area. For the velocity distribution of particles at the outlet of the nozzle in gas jets or jets at the macroscopic scale, although Anderson, Nathanson, et al. have done some theoretical and experimental measurement work, but mostly focused on the axial direction of the nozzle combined with one-dimensional theoretical results. Discussion, lacking a more complete theoretical description ~\cite{Nathanson2004, Anderson2004}. In the simulation process of nozzles on the micro-nano scale, it was found that the adjustment of the geometric parameters would cause complex changes in the outlet flow velocity distribution. Based on the above considerations, this paper approximates the divergence of the surface source by superimposing a disturbance velocity distribution when the particle stream enters the simulation area. In the following simulation results, the perturbation velocity uses the same probability density distribution function as the aforementioned point particle source. The perturbation velocity distribution parameter is set to $\overline{\epsilon} = 0.004 eV$. The laser beam waist position $z_0$ was adjusted to $20 \mu m$, and the other parameters were the same as those of the spot particle source simulation.

\begin{figure}[!htb]\centering
\includegraphics[width=\linewidth]{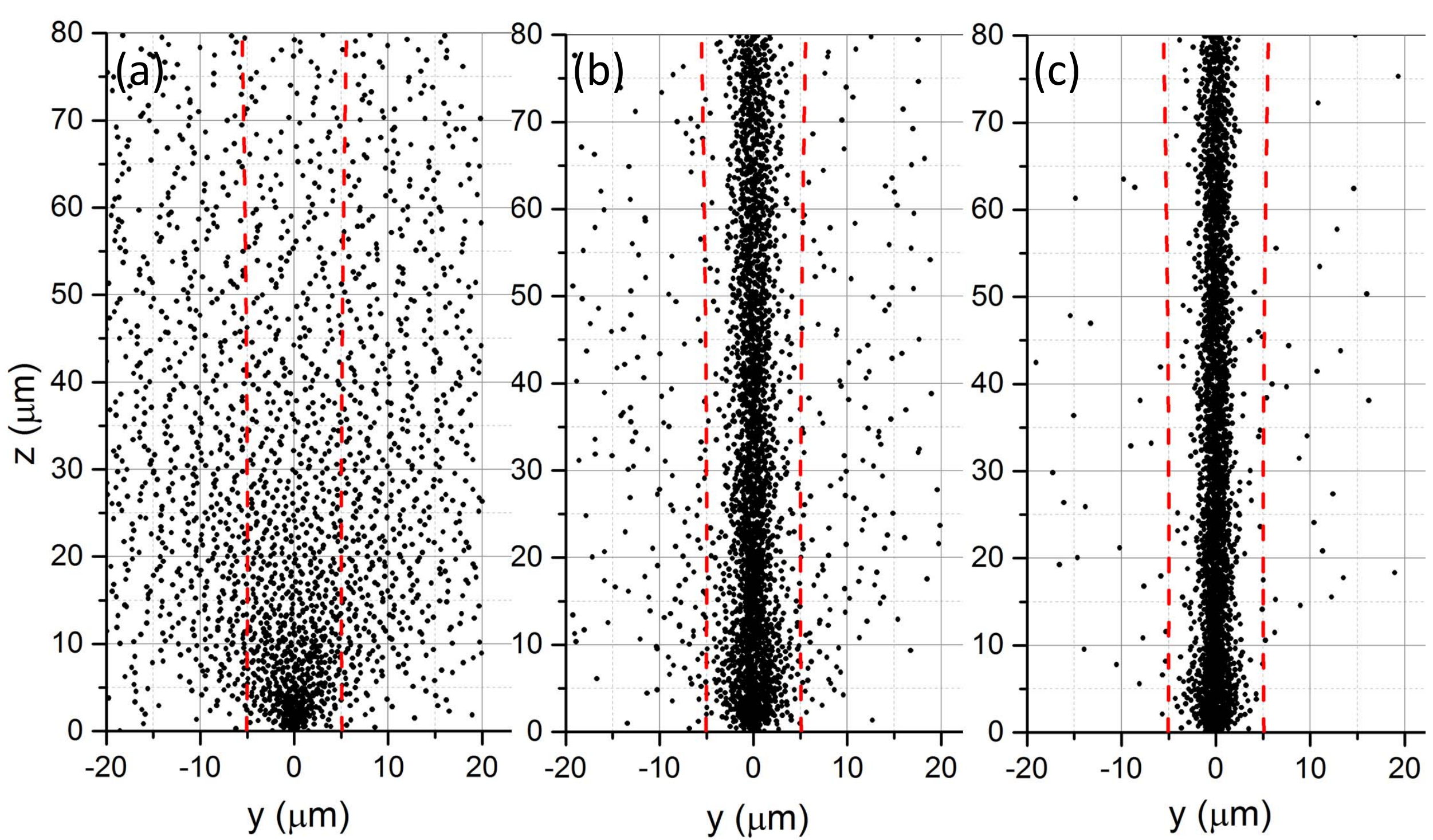}
\caption{screenshots of the spatial distribution of particle jets at different powers. (a). $P = 0.0W$. (b). $P = 0.5W$. (c). $P = 1.0W$.}
\label{fig:fig5}
\end{figure}

In order to reduce the amount of calculations during simulation, this paper refers to the particle injection method by Moseler et al. ~\cite{Moseler2000}. First, simulate a group of $300,000$ particles randomly moving in a cylindrical tube with a diameter of $1\mu m \times 20\mu m$ as the liquid column, and fix the relative positions of the particles. The liquid column was injected at a fixed speed of $v_z=0.1 m/s$, the relative positions of the particles entering the simulation region were released and the disturbance velocity was superimposed on the $v_z$ basis.

\begin{figure}[!htb]\centering
\includegraphics[height=6cm]{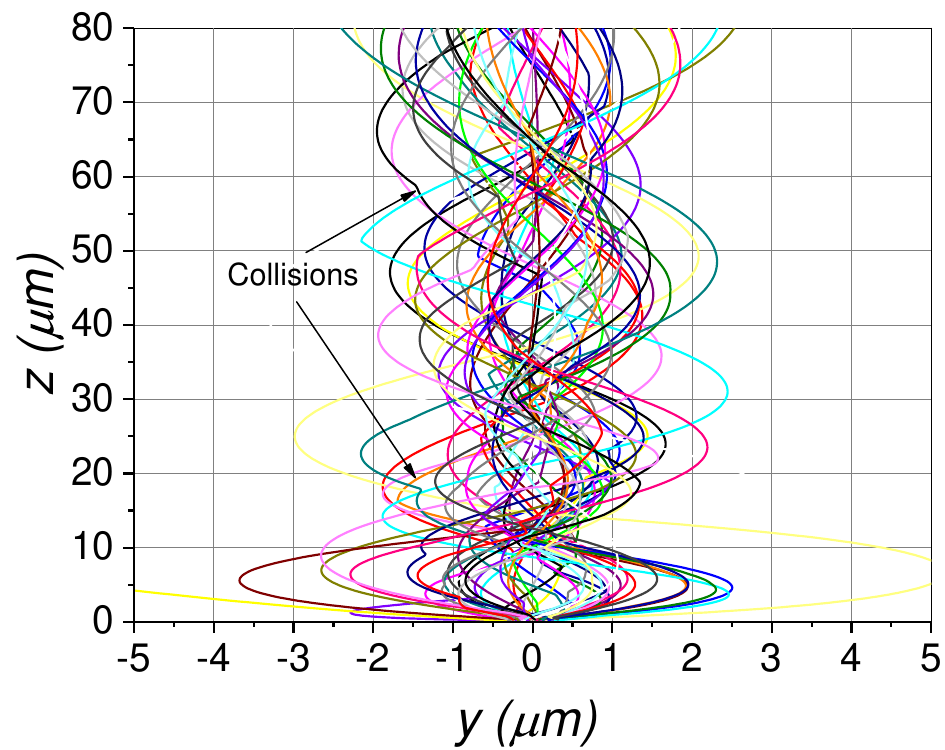}
\caption{Trajectories of $50$ adjacently numbered particles at $P = 1.0W$.}
\label{fig:fig6}
\end{figure}

\begin{figure}[!htb]\centering
\includegraphics[height=6cm]{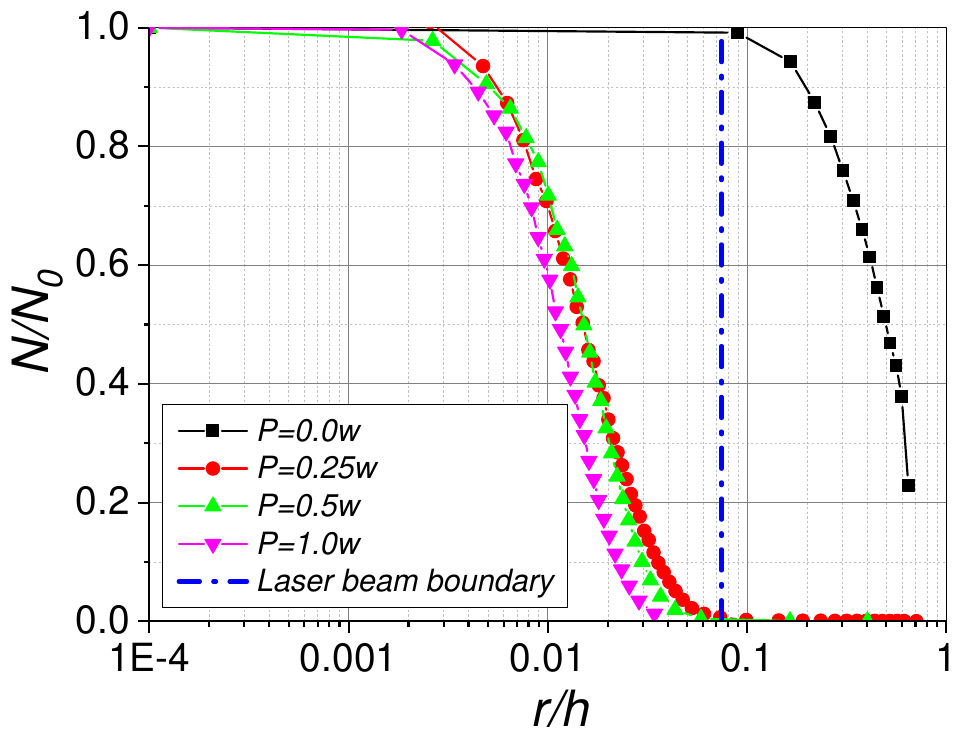}
\caption{Non-dimensional cumulative distributions of particle jeta at $h = 80 \mu m$ counting plane at different powers.}
\label{fig:fig7}
\end{figure}

\begin{figure}[!htb]\centering
\includegraphics[height=6cm]{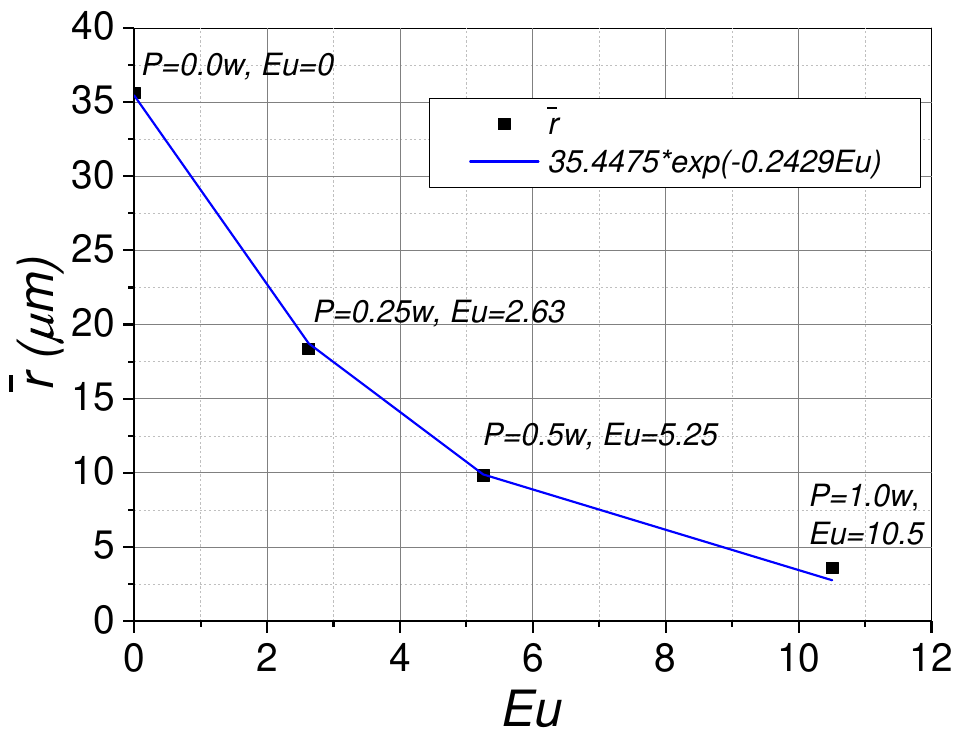}
\caption{Dispersive distribution variations on $h = 80 \mu m$ counting plane according to changes of electro-Euler number.}
\label{fig:fig8}
\end{figure}

Fig. 5 is the X-directional projection of the spatial distribution of a particle flow at a time with different power ($P=0.0 W, 0.5 W, 1.0 W$) Laser band. The dotted line in the figure is shown as the boundary of the laser beam. It can be seen that the spatial distribution density of particles is obviously changed. With the increase of power, the particle density in fig. 5 is gradually reduced. This is consistent with the conclusion of the Huang Kai, and increasing the laser power (electro-Euler number) can make the system to form a stable jet. At the same time, there is no obvious change in the diameter of the jet in the simulated region, that is, the instability phenomenon in the traditional jet theory. This is consistent with the conclusion that the Huang Kai laser will suppress the long-wave and the short wave instability. Fig. 6 is the flight trajectory of $50$ adjacent numbered particles in $P=1.0 W$. In this paper, the collision is judged by calculating the angle change of the flight trajectory in two adjacent time steps of the particle after the simulation. In the simulation of Fig. 5 (c) $P=1.0 W$, the average collision number of particles in the simulated domain is $12.2$ times, the $10^\circ $ time is 5.66 times, and the average collision frequency is about $15.3 \times {10^3}Hz$ and $7.08 \times {10^3}Hz$ respectively $5^\circ $ (The collision frequency of the molecules in the air is ${10^9}{10^{10}}{\rm{Hz}}$ level, i.e. $10^2$ order of magnitude for velocity and $10^{-7}$  order of magnitude for molecular free path). The particles that fly out of the beam range are about $2.36\%$ of the total grains.

On the $h=80\mu m$ plane, statistics of cumulative distribution densities at different powers are obtained in FIG. 7 , and it is found that the distribution of particles in the beam range under the simulated power is not significantly different. Use equation (18) to define the mean square radius $\bar r$ to represent the degree of dispersion of the particle position in the entire plane and observe the laser beam convergence effect. It can be seen from the curve fitted in the figure that with the linear growth of the electro-Euler number, the dispersion degree tends to decrease similarly to a negative index.

\section{Conclusion}

In this paper, the dynamic simulation of the superposition of the laser light field by the nanoparticle flow generated by the point source and the jet is realized based on the Lammps software. The stability of the nanoparticle jet under laser irradiation is analyzed by theoretical analysis. Comparing the point source simulation results with the theoretical distribution of classical vacuum evaporation deposition, the enhancement of the laser power $P$ can increase the concentration of the particle distribution. It is initially proved that the laser can effectively influence the particle motion trajectory, thereby improving the deposition and distribution of the nano-scale particles.

The theoretical analysis of the stability of photonic nanoparticle jets shows that increasing the electro Euler number $Eu$ can improve the stability of the particle jet to form a stable jet. At the same time, it is found that the laser suppresses the long-wave and short-wave stability of the particle jet at the same time. The simulation results of particle jets show that increasing electro Euler number $Eu$ by increasing the laser power $P$ can indeed constrain the particle formation jet more effectively, and the mean square radius decreases with the increase of the laser power, indicating that the concentration of particles also follows. The increase in electro-Euler number $Eu$ has increased, demonstrating a trend of stability that is consistent with theoretical predictions.

\begin{acknowledgements}
 This work was supported by National Natural Science Foundation of China (11172163), and also partially by the Shanghai Natural Science Foundation (15ZR1416400) and the National Key Research and Development Program (2017YFB0404503).
\end{acknowledgements}


\phantomsection
\bibliographystyle{apsrev}
\bibliography{lightbeam}
\listoffigures
\tableofcontents
\end{document}